\documentclass[lettersize,journal]{IEEEtran}
\usepackage{amsmath,amsfonts}
\usepackage{algorithm}
\usepackage{array}
\usepackage[caption=false,font=normalsize,labelfont=sf,textfont=sf]{subfig}
\usepackage{textcomp}
\usepackage{stfloats}
\usepackage{url}
\usepackage{verbatim}
\usepackage{graphicx}
\usepackage{cite}
\usepackage[T1]{fontenc}
\usepackage{makecell}
\usepackage{arydshln} 
\usepackage{booktabs}
\usepackage{tabularx}
\usepackage{algpseudocode}%
\usepackage{listings}%
\usepackage{colortbl}
\usepackage{diagbox}
\usepackage{multirow}
\usepackage{times}
\usepackage{multicol}
\usepackage{manyfoot}%
\usepackage{booktabs}%
\usepackage{xcolor}%
\usepackage{tabulary}
\usepackage{color}
\begin{document}

\definecolor{reb}{RGB}{3, 2, 164}
\newcommand{\reb}[1]{\textcolor{reb}{#1}}

\title{Reply with Sticker: New Dataset and Model for Sticker Retrieval}

\author{Bin~Liang$^\dagger$,
        Bingbing~Wang$^\dagger$,
        Zhixin~Bai,
        Qiwei~Lang,
        Mingwei~Sun,
        Kaiheng~Hou,
        Lanjun~Zhou,
        Ruifeng~Xu$^\star$,~\IEEEmembership{Member,~IEEE},
        and~Kam-Fai~Wong$^\star$,~\IEEEmembership{Senior Member,~IEEE}
\IEEEcompsocitemizethanks{
% \IEEEcompsocthanksitem 
\IEEEcompsocthanksitem B. Liang, B. Wang and R. Xu are with the School of Computer Science and Technology, Harbin Institute of Technology (Shenzhen), Shenzhen, China, and also with the Guangdong Provincial Key Laboratory of Novel Security Intelligence Technologies, Shenzhen, China.\protect\\
E-mail: \{bin.liang, bingbing.wang\}@stu.hit.edu.cn, xuruifeng@hit.edu.cn.
\IEEEcompsocthanksitem K. Wong is with the Department of Systems Engineering and Engineering Management, The Chinese University of Hong Kong, Hong Kong, China.\protect\\
E-mail: kfwong@se.cuhk.edu.hk;
\IEEEcompsocthanksitem Z. Bai is with the  School of Computer Science and Technology, Harbin Institute of Technology, Harbin, China.\protect\\
E-mail: baizhixin@stu.hit.edu.cn;
\IEEEcompsocthanksitem Q. Lang, M. Sun, and K. Hou are with the School of Computer Science and Technology, Harbin Institute of Technology (Shenzhen), Shenzhen, China.
\protect\\
E-mail: langqw5520@mails.jlu.edu.cn, hitsunmingwei@gmail.com, houkaiheng@stu.hit.edu.cn;
\IEEEcompsocthanksitem L. Zhou is with the School of Computer Science and Artificial Intelligence, Foshan University, Foshan, China.\protect\\
E-mail: bluejade.zhou@gmail.com;
\IEEEcompsocthanksitem $^\dagger$ Equal contribution.~~$^\star$ Corresponding Authors.
}

}

\markboth{Journal of \LaTeX\ Class Files,~Vol.~14, No.~8, August~2021}%
{Shell \MakeLowercase{\textit{et al.}}: A Sample Article Using IEEEtran.cls for IEEE Journals}

\IEEEpubid{0000--0000/00\$00.00~\copyright~2021 IEEE}

\maketitle
\begin{abstract}
Using stickers in online chatting is very prevalent on social media platforms, where the stickers used in the conversation can express someone's intention/emotion/attitude in a vivid, tactful, and intuitive way. 
Existing sticker retrieval research typically retrieves stickers based on context and the current utterance delivered by the user. That is, the stickers serve as a supplement to the current utterance. However, in the real-world scenario, using stickers to express what we want to say rather than as a supplement to our words only is also important. Therefore, in this paper, we create a new dataset for sticker retrieval in conversation, called \textbf{StickerInt}, where stickers are used to reply to previous conversations or supplement our words\footnote{We believe that the release of this dataset will provide a more complete paradigm than existing work for the research of sticker retrieval in the open-domain online conversation.}.
Based on the created dataset, we present a simple yet effective framework for sticker retrieval in conversation based on the learning of intention and the cross-modal relationships between conversation context and stickers, coined as \textbf{Int-RA}. Specifically, we first devise a knowledge-enhanced intention predictor to introduce the intention information into the conversation representations. Subsequently, a relation-aware sticker selector is devised to retrieve the response sticker via cross-modal relationships.
Extensive experiments on two datasets show that the proposed model achieves state-of-the-art performance and generalization capability in sticker retrieval\footnote{The dataset and source code of this work are released at \url{https://github.com/HITSZ-HLT/Int-RA}.}.
\end{abstract}


\section{Introduction}
With the rise of instant messaging applications, online chatting has become an essential part of daily life \cite{zhang2024stickerconv}. Stickers, as visual elements on social platforms, bring a dynamic and multifaceted dimension to conversations. Previous research on stickers has largely concentrated on sentiment analysis \cite{2-liu2022ser30k,3-zhao2023sticker820k,6-ge2022towards}. Due to their visual appeal, stickers uniquely contribute to fostering a lively and innovative conversational environment \cite{nilasari2018sticker,albar2018chat}. Therefore, incorporating automatic sticker replies based on previous conversations into dialogue systems can significantly enhance the engagement and liveliness of interactions.

\begin{figure}[!t]
  \centering
  \includegraphics[width=\linewidth]{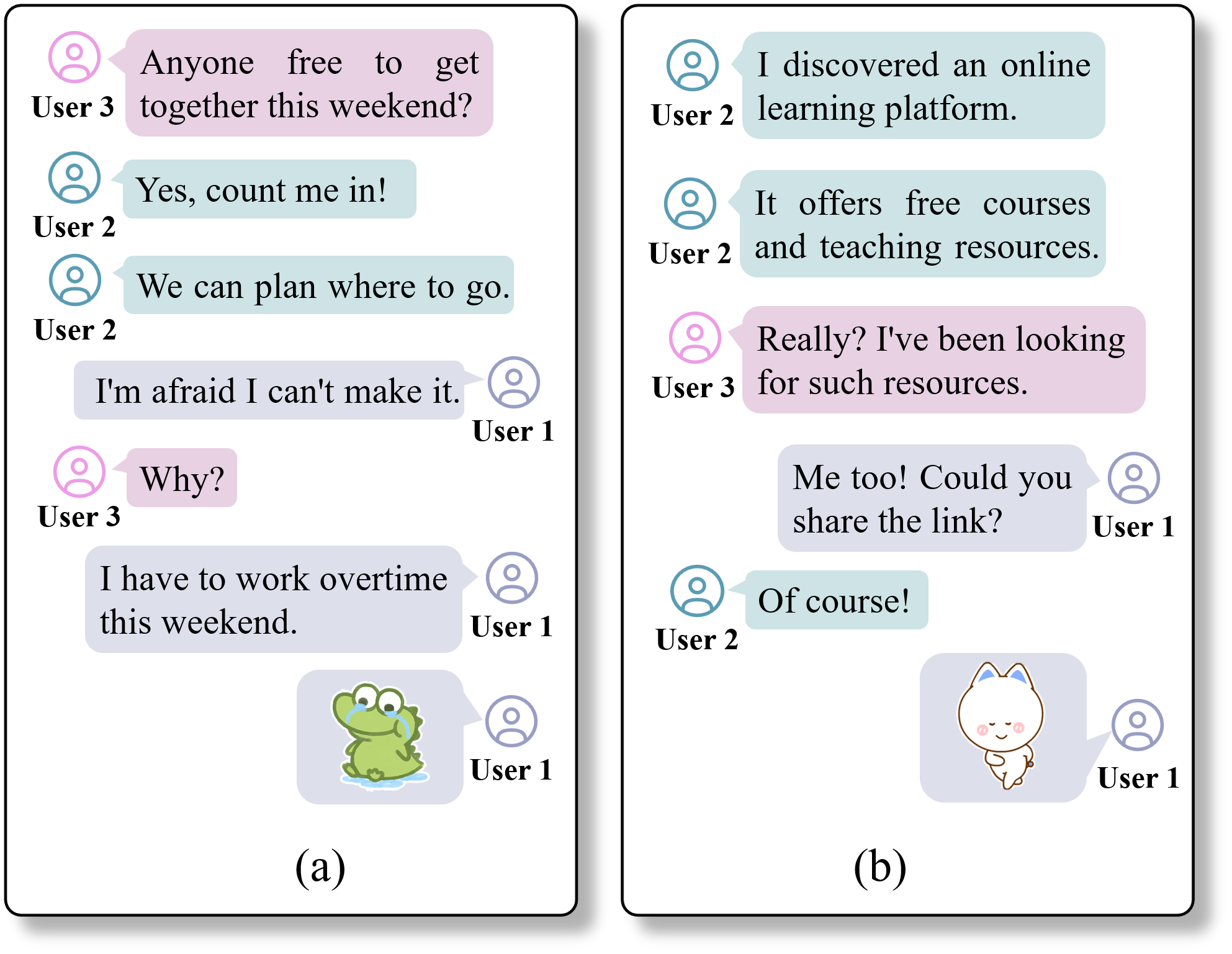}
  \caption{Two examples of stickers used in an online conversation.}
  \label{f-example}
  % \vspace{-20pt}
\end{figure}

Recent research endeavors of sticker retrieval have been dedicated to using stickers to supplement the current response, in order to strengthen the expression of emotion, attitude, or opinion~\cite{4-fei2021towards,5-gao2020learning}. However, in the real-world scenario, we may also use stickers to reply to the previous conversation directly, rather than merely supplementing our words with stickers. 
We refer to the former as the Supplementary Response (SR) scenario, where stickers are used alongside the speaker’s own utterance, and to the latter as the Direct Response (DR) scenario, where stickers function as independent replies without accompanying text.
One expected scenario of the sticker retrieval task is that suitable stickers can be retrieved for replies whether or not one has made a textual response. To illustrate our idea, we give examples shown in Figure~\ref{f-example}. Figure~\ref{f-example} (a) shows the SR scenario, and Figure~\ref{f-example} (b) depicts the DR scenario. Therefore, in our work, we create a new sticker retrieval dataset to cover these two scenarios, called \textbf{StickerInt}, which is a comprehensive consideration for using stickers in social media conversation. Our \textbf{StickerInt} dataset contains 1,578 Chinese conversations with 12,644 utterances.

\IEEEpubidadjcol
Based on our \textbf{StickerInt} dataset, we propose a pipeline framework, called \textbf{Int-RA}, which comprises a knowledge-enhanced intention predictor and a relation-aware selector to retrieve the sticker for responding to the conversation. Specifically, for the text modality, we feed the conversation context into the BART version of COMET \cite{bosselut2019comet} to generate commonsense inferences of five relations (\textit{xIntent, xNeed, xWant, xReact, xEffect}) based on the commonsense knowledge base ATOMIC$_{20}^{20}$ \cite{hwang2021comet}, which are concatenated with conversation contexts through a text encoder to derive the textual representation. Here, the textual representation is fed into a classifier to infer the intention of the user towards the conversation with the help of supplemental tags, deriving intention-fused textual representation. For visual modality, we feed the sticker into the visual encoder to get the visual representation of the sticker's regions. Further, to better learn the properties of stickers, we define four attributes \textit{gesture}, \textit{posture}, \textit{facial expression}, and \textit{verbal}, which are used as prompts in the multi-modal large language model (MLLM) to derive attribute-aware sticker descriptions. Afterward, we use cross-modal attention to learn the relationship between the visual representation of regions and descriptions for each sticker, deriving the relation-aware visual representation. Finally, we calculate the similarity between the intention-fused textual representation and each relation-aware visual representation to determine the result of sticker retrieval.

Our main contributions can be summarized as follows:
\begin{itemize}
    \item To facilitate the research of sticker retrieval, we create \textbf{StickerInt}, a novel and more comprehensive sticker retrieval dataset for sticker usage in conversation. Further, our \textbf{StickerInt} dataset provides an intention tag for each sticker towards the conversation, aiming at empowering models with more in-depth learning on sticker retrieval. 
    \item We propose a novel pipeline framework for sticker retrieval, in which a knowledge-enhanced intention predictor and relation-aware selector are devised to leverage the intention information and fine-grained sticker attributes to retrieve stickers for response.
    \item Experiments conducted both on our created \textbf{StickerInt} and a public dataset demonstrate that our proposed framework outperforms the baseline models.
\end{itemize}

\section{Related Work}
\subsection{Sticker Dataset}
Sticker analysis has attracted more and more attention in recent years. Numerous works fix their eyes on sticker-based multi-modal sentiment analysis and have proposed a wide variety of multi-modal sentiment analysis datasets \cite{2-liu2022ser30k,3-zhao2023sticker820k}.
However, stickers, being visual images, demonstrate significant effectiveness within conversations beyond mere research into the stickers themselves. Consequently, several researchers have shifted their focus towards retrieving stickers based on the context of the conversation.
\cite{4-fei2021towards} introduced a novel task termed meme incorporated open-domain conversation and further built a large dataset of 45k Chinese conversations with 606k utterances. 
A Chinese sticker-based multi-modal dataset for the sentiment analysis task (CSMSA) \cite{6-ge2022towards} was presented by collecting from eight public open chat groups with 28k text-sticker pairs and 1.5k annotated data. 
In these datasets, the stickers serve as a supplement to the current utterance. Nevertheless, in real-world scenarios, stickers are used not just as supplements to words but also as a means to directly express the user's intentions. Therefore, we create a comprehensive dataset for sticker retrieval in conversation.

\subsection{Image Retrieval}
Image retrieval aims to retrieve images that match a given query from a large collection of images. Early method \cite{faghri2017vse++} mainly relied on global feature-based image-text matching.
Further, \cite{li2019visual} proposed a Visual Semantic Reasoning Network (VSRN), which generates global features with regional semantic associations through a Graph Convolutional Network (GCN).
To address the false negative problem in existing image-text matching benchmarks, \cite{li2023integrating} proposed an ITM framework integrating Language Guidance. 
In terms of improving retrieval efficiency and accuracy, \cite{zhuang2023towards} proposed a self-supervised fine-grained alignment module called SelfAlign. 
\cite{wu2023feature} proposed a feature-first approach to advance image-text retrieval by improving visual features. This method generates more comprehensive and robust visual features through a two-step interaction model and a multi-attention pooling module.
Recently, with the development of large-scale pretraining models like CLIP \cite{radford2021learning} and ALIGN \cite{li2021align}, significant performance improvements are achieved in image-text matching tasks by pretraining on massive vision-language data and then fine-tuning for specific downstream tasks.
Different from the image retrieval methods, we focus on retrieving the image based on the conversation not just the text.

\subsection{Multi-modal Conversation}
Several multi-modal studies have sought to improve the efficacy of conversational agents by enriching textual expressions of generated dialog responses through associative vision scenes \cite{7-zang2021photochat, 8-yang2021open,9-chen2021learning,LLM-zhao2024narrativeplay}.
In contrast, sticker retrieval in the real-world  requires understanding the semantic expression and intention of the context and the sticker image. Several works specifically paid attention to sticker features \cite{10-laddha2020understanding, 6-ge2022towards}.
\cite{10-laddha2020understanding} proposed a real-time system for sticker recommendation, which decomposes the sticker recommendation task into two steps, including message prediction that the user is likely to send and an appropriate sticker substitution. 
Nowadays, using stickers for replies has become commonplace in social media interactions.
There has been a growing number of works on sticker retrieval, which assists users in selecting the appropriate sticker for a response.
\cite{5-gao2020learning} proposed a sticker response selector model that employs a convolution-based sticker image encoder and a self-attention-based multi-turn dialog encoder to obtain the representation of stickers and utterances. Then, a deep interaction network is designed for obtaining deep matching, and a fusion network is employed to output the final matching score.
\cite{4-fei2021towards} presented a Meme incorporated Open-domain Dialogue (MOD) task and utilized pooling all sub-tasks like text generation and internet meme prediction into a unified sequence generation procedure to solve it. 
To tackle three challenges of inherent multimodality, significant inter-series variations, and complex multi-modal sentiment fusion. \cite{6-ge2022towards} integrated both the sticker text and series tag to holistically model sticker sentiment. They employed a flexible masked attention mechanism to selectively extract the most pertinent information crucial for the current sentiment assessment.
These methods, which directly match conversation context and stickers, overlook the expressive role of stickers in conversation, where emotions and intentions are conveyed in a visually engaging manner.

\section{StickerInt Dataset}
This section introduces our new dataset \textbf{StickerInt} for sticker retrieval in detail. Specifically, Section~\ref{sec-data-construction} presents the construction of our \textbf{StickerInt} dataset, Section~\ref{sec-data-annotation} introduces the annotation process of \textbf{StickerInt}, and Section~\ref{sec-data-analysis} analyzes the created \textbf{StickerInt} by data statistics.

\subsection{Data Construction}
\label{sec-data-construction}
We collected our dataset from a widely used social platform (WeChat\footnote{\url{https://weixin.qq.com/}}), which boasts a vast number of conversations with stickers accessible for both individual and group chats. We select four open chat groups with active participants and collect their conversations. Among them, each group engages in open-domain online conversation, making the use of stickers more diverse. Note that we eliminate extraneous image elements like screenshots and photos. 

We formulate stringent guidelines and policies for data preprocessing. 
Specifically, we follow established best practices adopted in prior work such as \cite{gao2020learning} and \cite{privacy-melo2024sticker}, and implement a two-stage anonymization pipeline. (1) In the first stage, we conduct \textbf{automated removal and replacement of direct identifiers}. All direct user-related information, including user IDs, phone numbers, nicknames, and profile metadata, is systematically removed or replaced with pseudonyms. To preserve the continuity and coherence of multi-turn conversations, we assign consistent anonymized tokens such as "User\_1" and "User\_2" to represent different participants. Usernames are replaced rather than deleted to retain clarity in speaker attribution while safeguarding privacy.
(2) In the second stage, we perform a \textbf{manual review to identify and anonymize any remaining indirect identifiers}. These may include references to personal occupations, geographic locations, interpersonal relationships, or infrequent linguistic patterns that could potentially lead to re-identification. For such instances, we apply additional anonymization strategies such as data generalization to reduce specificity and data suppression to remove sensitive elements entirely.

In addition to protecting user privacy, we remove any content that contains inappropriate, offensive, or abusive language to comply with ethical research standards. To ensure contextual integrity, we segment the chat history into distinct and coherent conversations. Each sticker in the dataset is aligned with its surrounding conversational context, guaranteeing that it is situated within a meaningful and self-contained dialogue.

\begin{table}[!t]
\centering
\caption{Statistics of StickerInt dataset. con. = conversation, Avg. = average. SR and DR represent Supplementary Response and Direct Response Scenarios.}
\resizebox{\linewidth}{!}{
\begin{tabular}{l|ccc}
\hline  
 \textbf{Dataset Statistics} & \textbf{Train} & \textbf{Valid} & \textbf{Test} \\ \hline
\# con. &1,269& 155	& 154 \\
\# SR scenario & 816&102 &99\\
\# DR scenario &453&53&55\\
\# utterances	&8,745 &1,105&1,216 \\
\# tokens	&47,895&5,950&5,778 \\
\# stickers	& 774 &127&124\\ 
\# users	& 48 &41&40\\ \hline
Avg. utterances in a con. &6.89&7.13&7.90\\
Avg. users in a con. &2.91&2.71&2.78\\
Avg. tokens in an utterance &5.48 &5.39 &4.75\\
\hline
\end{tabular}
}
\label{tab1}
\end{table}

\subsection{Data Annotation}
\label{sec-data-annotation}
% 补充style and form

We recruited 5 experienced researchers with over 3 years of research experience in the field of multi-modal learning as annotators to check and label the golden sticker for each conversation, aiming at eliminating the impact of noise stickers on the research of sticker retrieval.

In addition, recognizing the sentiment, emotion, or feeling can help us better select stickers for replies. Therefore, considering the diversity and complexity of intentional expression in conversation, we get inspiration from \cite{aman2008using} and use GoEmotions \cite{demszky2020goemotions} to request annotators to supplement an intention tag for each sticker towards the conversation.
We enlist the expertise of five annotators to label both the coarse and fine-grained intention of stickers based on the dialogues. Given that the outcome of the annotation process is closely tied to the annotators' subjective judgment, each annotator is initially tasked with annotating 500 sticker samples. This initial task aids them in comprehending the annotation process and understanding the nuances of intention, thereby enhancing both the efficiency and accuracy of the subsequent labeling endeavors.

\begin{figure}[!t]
  \centering
  \includegraphics[width=\linewidth]{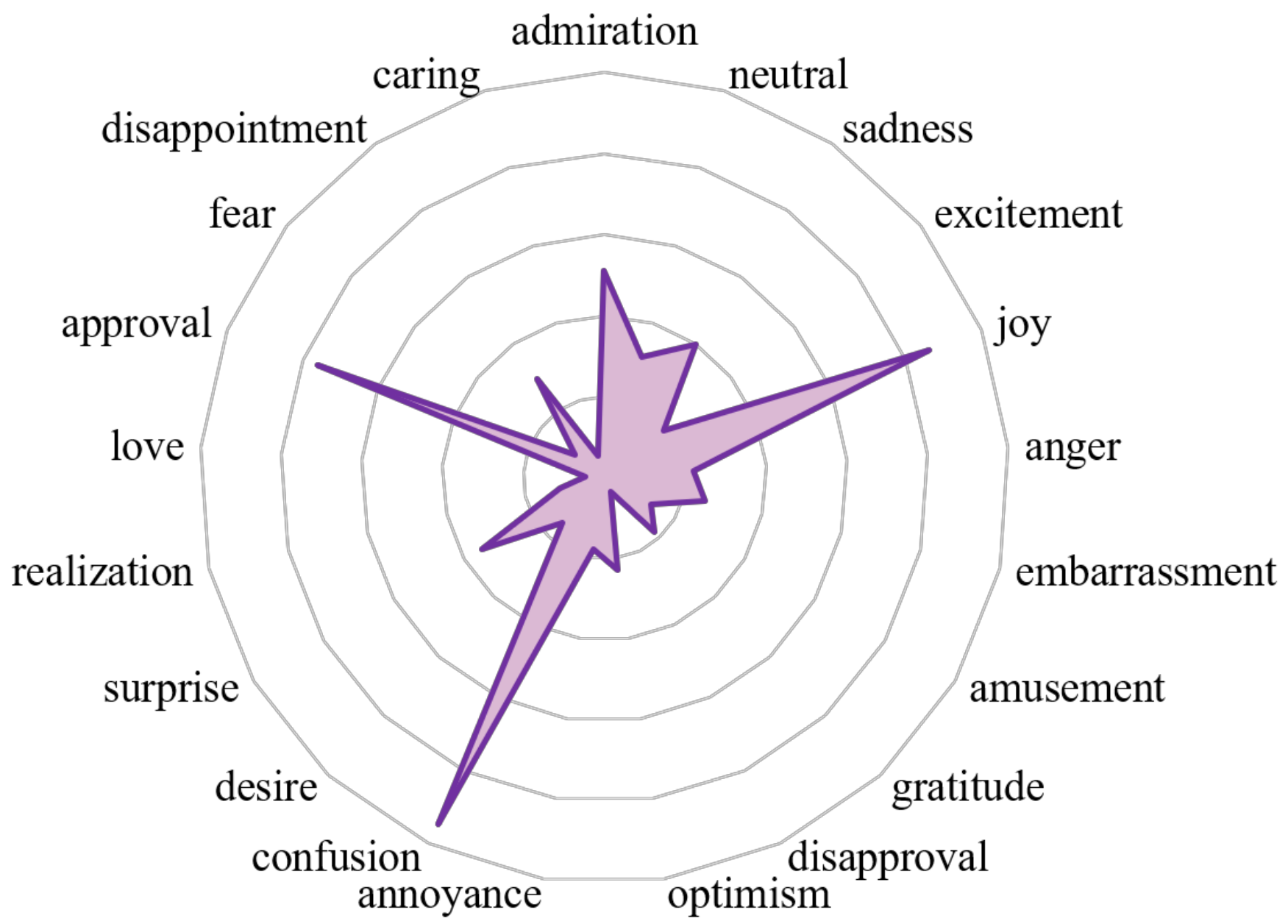}
  \caption{Visualization depicting the distribution of intention labels.}
  \label{f2}
\end{figure}

\begin{figure}[!t]
  \centering
  \includegraphics[width=0.8\linewidth]{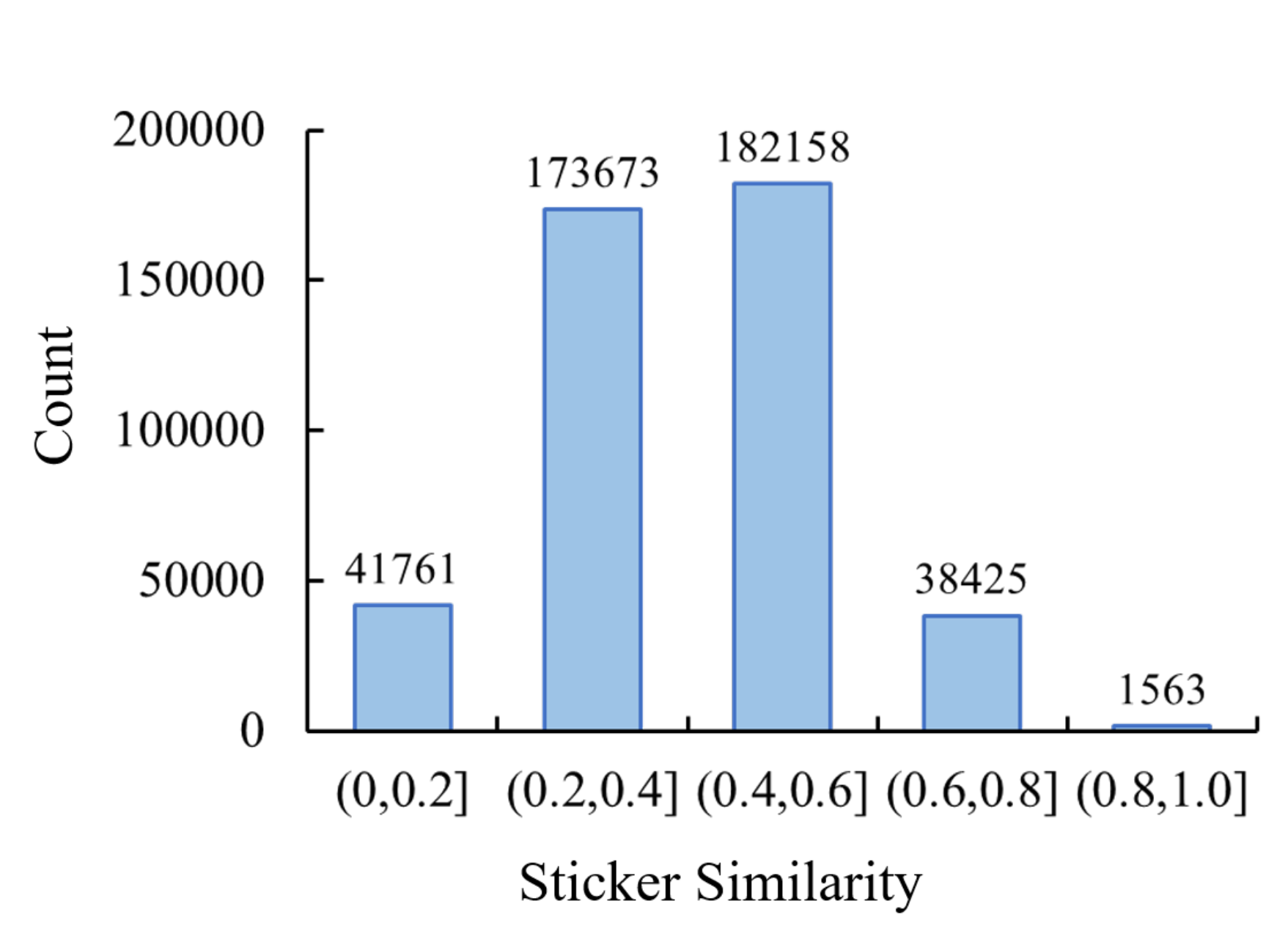}
  \caption{Similarity distribution among all stickers in our dataset. The x-axis represents different ranges of sticker similarity, and the y-axis indicates the count of SSIM calculations.}
 \label{similarity}
\end{figure}

\subsection{Dataset Analysis}
\label{sec-data-analysis}
The detailed statistics of our dataset are shown in Table \ref{tab1}. In total, there are 1,578 conversations that contain 12,644 utterances, 59,623 tokens, and 1,025 stickers. Of these stickers, 702 are associated with textual content. The proposed dataset is collected from conversations with Chinese users on WeChat, and the dialogues are predominantly in Chinese. Notably, 841 utterances include English words such as “OK”, “QQ”, and “NLP”, which are naturally embedded in Chinese expressions, reflecting the way people commonly communicate in daily life. 
Each conversation includes 8.01 utterances on average. 
The average number of users who participate in a conversation is 2.80.
Among all conversations in our dataset, 1,017 belong to the Supplementary Response scenario, and 561 fall under the Direct Response scenario (sticker as a direct reply to previous conversation turns).
The average number of tokens in an utterance is 5.21.
Furthermore, we also visualize the distribution of intention tags in Figure \ref{f2}. We can see that the proportion of stickers used in different labels varies, which also demonstrates the diversity of stickers' intention expressions in our dataset.

\subsection{Sticker Similarity}
Stickers always share a similar style or contain the same cartoon characters. Intuitively, the more similar the stickers are, the more difficult it is to select the correct sticker from the sticker set. 
In other words, the similarity between stickers determines the difficulty of the sticker retrieval task. To investigate the difficulty of this task, we calculate the average similarity of all the stickers in our sticker set by the Structural Similarity Index (SSIM) metric \cite{wang2004image}.
We calculate the SSIM between each two stickers and normalize it into [0, 1]. The similarity distribution among our sticker data is shown in Figure \ref{similarity}, where the average similarity is 0.4016.

\begin{figure*}[!t]
  \centering
  \includegraphics[width=\linewidth]{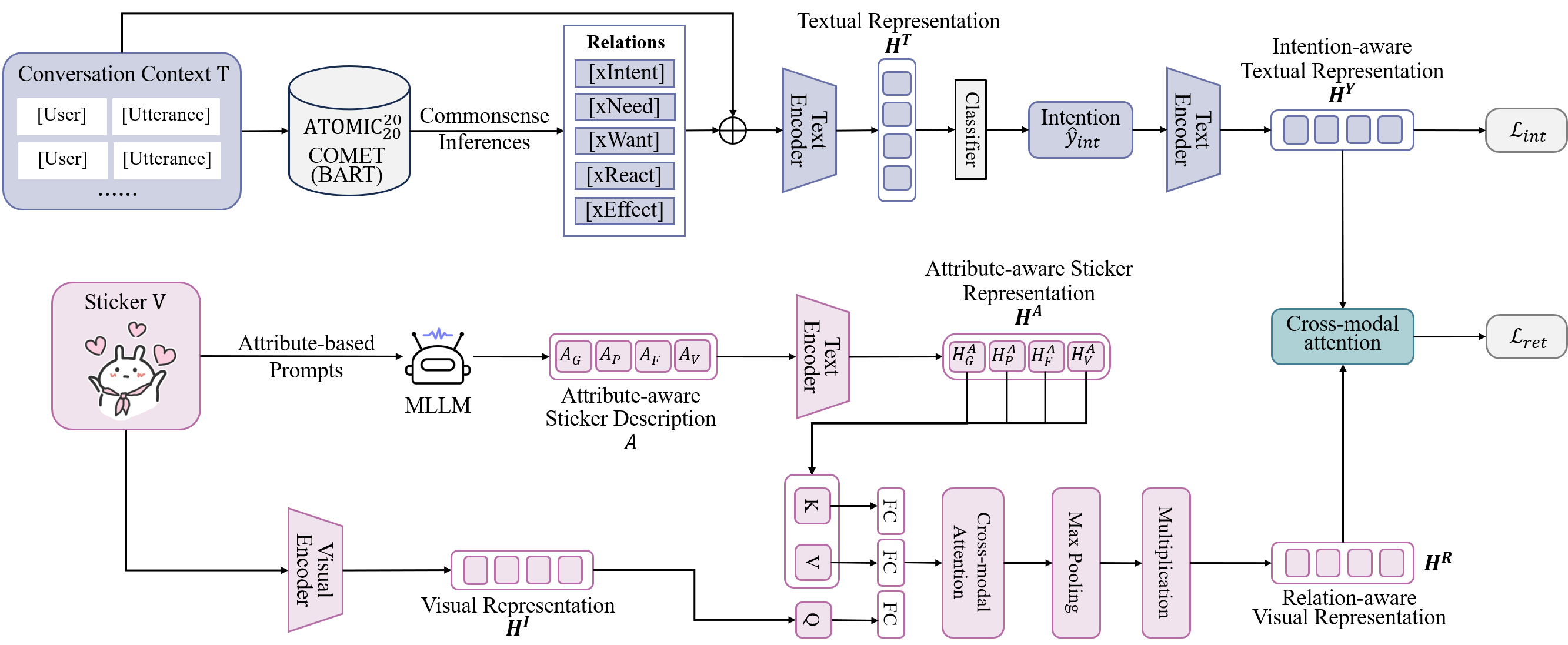}
  \caption{Illustration of the proposed Int-RA framework comprising intention-fused conversation context representation, attribute-aware sticker representation,  and relation-aware sticker selector.}
  \label{f3}
\end{figure*}

\section{Methodology}
%\section{Task Definition}
In this section, we introduce our proposed \textbf{Int-RA} framework for sticker retrieval in detail. Assume that there is a conversation context $T=\{t_1,...,t_{N_t},s_j\}$ and a sticker set $V=\{v_1,...v_{N_v}\}$, where $t_i=(s_i, u_i)$, $s_i$ and $u_i$ represent the $i$-th user and utterance. $N_{t}$ and $N_{v}$ represent the number of utterances and stickers, respectively. 
The sticker retrieval task aims to select a suitable sticker from the sticker set $V$ based on the conversation context $T$ for the user $s_j$.
In the $i$-th utterance $u_i=\{w_1^i,...,w^i_{N_w^i}\}$, $w_j^i$ represents the $j$-th word in $u_i$, and $N_x^i$ represents the total number of words in $u_i$ utterance. 
Additionally, our dataset presents an annotated intention tag $y_{int}$ as supplemental information in each conversation $T$.
In the dialog context $T$, $s_i$ represents a sticker image with a binary label $y_i$, indicating whether $s_i$ is an appropriate response for $T$. Each sticker with the intention label $m_i$ indicates the intention of the speaker to use the sticker to respond. Our goal is to learn a ranking model that can produce the correct ranking for each dialog.

Therefore, by leveraging the intention tags, we propose a pipeline framework (\textbf{Int-RA}) to deal with the sticker retrieval task. The architecture of our \textbf{Int-RA} is illustrated in Figure~\ref{f3}, which mainly comprises three components: 
1) \textit{\textbf{Intention-fused Conversation Context Representation}}, which introduces intention information to the learning of conversation context representation based on commonsense;
2) \textit{\textbf{Attribute-aware Sticker Representation}}, which uses the multi-modal large language model (MLLM) to derive the representation of a sticker based on the attribute-based prompts;
3) \textit{\textbf{Relation-aware Sticker Selector}}, which facilitates the retrieval of the relationships between conversation context and stickers by cross-modal attention.

\subsection{Intention-fused Conversation Context Representation}
For online chatting, we generally use stickers to express our sentiments, status, feelings, etc. Therefore, to explore the potential sentiment, status, or feeling expressed by the user towards a conversation context for sticker retrieval, we devise a knowledge-enhanced intention predictor, aiming to introduce intention information into the conversation context representation. 

Inspired by recent works \cite{sabour2022cem, qian2023harnessing}, we adopt the commonsense knowledge base $\text{ATOMIC}_{20}^{20}$ \cite{hwang2021comet}, which contains knowledge not readily available in pre-trained language models and can generate accurate and representative knowledge for unseen entities and events.
Specifically, we utilize the BART version of COMET \cite{hwang2021comet} trained on this knowledge base to generate commonsense inferences of five relations including \textit{xIntent, xNeed, xWant, xEffect, xReact}.
\begin{equation}
    C_r = \text{COMET}(T,r),
\end{equation}
\begin{equation}
    C_{know} = \underset{R}{\oplus} C_r,
\end{equation}
where $r \in R$ denotes the relation type, $R \in \{\textit{xIntent, xNeed, xWant, xEffect, xReact}\}$. $\oplus$ is the concatenation operation. 
Then these five relations are concatenated with conversation contexts through a pre-trained Multi-lingual BERT (M-BERT) \cite{devlin2018bert} to derive the textual representation $\textbf{\textit{H}}^T$:
\begin{equation}
    \textbf{\textit{H}}^T = \text{M-BERT} (T \oplus C_{know}),
\end{equation}

Afterward, $\textbf{\textit{H}}^T$ is input to a softmax classifier to infer the user's intention $\hat y_{int}$ and optimized with cross-entropy loss, which can be described as:
\begin{equation}
   \rho = \text{softmax}(W\textbf{\textit{H}}^T+b),
\end{equation}
\begin{equation}
   \hat y_{int} = \text{argmax}(\rho),
\end{equation}
\begin{equation}
    \mathcal{L}_{int} = -\sum_{i=1}^n y_{int} \log \rho.
\end{equation}
where $W$ and $b$ are the weight matrix and bias term, respectively. $n$ represents the number of samples, and $y_{int}$ is the ground-truth intention tag. In this way, we can obtain a deeper understanding of the intention that may be expressed towards the conversation context, and introduce intention information into the textual representation.
The intention $\hat y_{int}$ is fed into M-BERT to obtain intention-fused textual representation $\textbf{\textit{H}}^Y=\text{M-BERT}(\hat y_{int})$.

\subsection{Attribute-aware Sticker Representation}

\begin{figure}[!t]
  \centering
  \includegraphics[width=\linewidth]{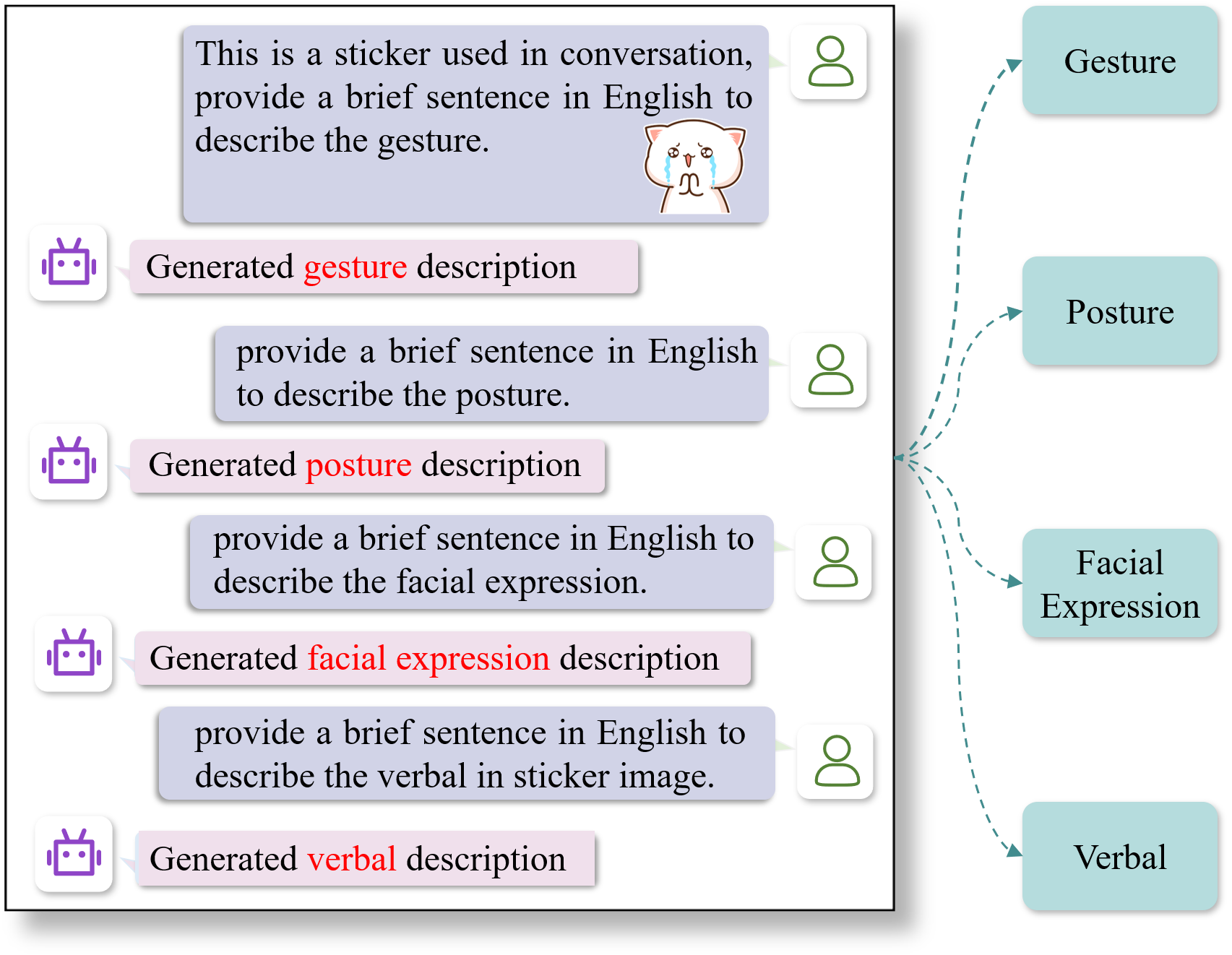}
  \caption{Overview of attribute-aware sticker description generation.}
  \label{attr}
\end{figure}

To capture key expressive features and reduce interference from irrelevant style variations in sticker learning, we adopt four representative attributes informed by prior research in visual communication and multimodal semantics. Studies have shown that gesture, facial expression, posture, and verbal elements are fundamental to the visual encoding of meaning. \cite{Theory-jessica2017analysis} highlights these visual signs as core denotative elements whose interpretation depends on contextual knowledge. \cite{Theory-afifah2020gesture} further demonstrates their frequent use in memes to convey intention and emotion. Guided by these findings, we define four attributes for sticker representation: gesture $L_G$, posture $L_P$, facial expression $L_F$, and verbal text $L_V$. Together, these features encompass both visual and textual cues, enabling more robust and fine-grained semantic modeling for sticker retrieval.
Based on this, we use Qwen-VL \cite{44-bai2023qwen} as the MLLM to produce attribute-aware sticker descriptions based on the above four attributes:
\begin{equation}
\begin{split}
        &\{A_G,A_P,A_F,A_V\}\\
    &=\text{MLLM}(\{L_G,L_P,L_F,L_V\}),
    \end{split}
\end{equation}

As demonstrated in Figure \ref{attr}, we use several turns of interactions, including the system prompt like “This is a sticker used in conversation, please provide several keywords to describe the gesture/posture/facial expression/verbal.” to simulate the utterance generation ability of MLLM. 
Then, each attribute-aware sticker description is transformed into an attribute-aware sticker representation using M-BERT: 
\begin{equation}
\begin{split}
        &\textbf{\textit{H}}^A = \{ \textbf{\textit{H}}^A_G,\textbf{\textit{H}}^A_P,\textbf{\textit{H}}^A_F,\textbf{\textit{H}}^A_V\}\\
    &=\text{M-BERT}(\{A_G,A_P,A_F,A_V\}).
    \end{split}
\end{equation}

Further, to learn the visual information of stickers, sticker $v_i$ first undergoes CLIP pre-trained ViT model \cite{dosovitskiy2020image} as a visual encoder to obtain visual representation $\textbf{\textit{H}}^I$.
\begin{equation}
    \textbf{\textit{H}}^I = \text{ViT}(v),
\end{equation}

Afterward, we adopt cross-modal attention between the visual representation and each attribute-aware sticker representation of the sticker to highlight the important regions in the sticker.
In detail, we use two fully connected layers $f_{vis}$ and $f_{des}$ to project the visual representation and description representation into the same dimension $d$: 
\begin{equation}
    \textbf{\textit{h}}_j^A = f_{des}(\textbf{\textit{H}}^A_j), \textbf{\textit{h}}^I = f_{vis}(\textbf{\textit{H}}^I),
\end{equation} 
where $\textbf{\textit{H}}^A_j, j \in \{G, P, F, V\}$.
$M_{j} \in \mathbb{R}$ indicates the relation between $\textbf{\textit{h}}^A_j$ and the visual representation $\textbf{\textit{h}}^I$, and can be expressed as:
\begin{equation}
    M_{j} =\text{softmax}(\frac{(\textbf{\textit{h}}_j^A W^Q)(\textbf{\textit{h}}^I W^K)^{\top}}{\sqrt{d_k}})(\textbf{\textit{h}}^I W^V).  
\end{equation}
where $W^Q \in \mathbb{R}^{d \times d_q}, W^K \in \mathbb{R}^{d \times d_k}, W^V \in \mathbb{R}^{d \times d_v}$ are randomly initialized projection matrices. We set $d_k, d_v,d_q = d/h$ for each of these parallel attention layers. $h$ is the number of heads in each multi-head attention layer.

Next, a max pooling operation is conducted on $M$, i.e., let $ \mathcal{M} = \text{max}(M_{j}) \in \mathbb{R}$ represent the relation score on the sticker by attribute-aware sticker descriptions. This attention learns to assign high weights to the important regions of the sticker that are closely related to each attribute-aware sticker description. 
We finally conduct a multiplication operation of each visual representation and relation score to obtain relation-aware visual representation $\textbf{\textit{H}}^R$ for the sticker.
\begin{equation}
    \textbf{\textit{H}}^R = \mathcal{M} \times \textbf{\textit{h}}^I.
\label{att_eqa}
\end{equation}

\subsection{Relation-aware Sticker Selector}
Ultimately, we leverage the relation-aware sticker representations to perform cross-modal retrieval.
We primarily implement the matching function using cosine similarity as cross-modal attention, which is defined as:
\begin{equation}
    CA = cos(\textbf{\textit{H}}^Y, \textbf{\textit{H}}^R).
\end{equation}

We optimize our method to minimize a learning objective: $\mathcal{L} = \lambda_1 \mathcal{L}_{ret} + \lambda_2 \mathcal{L}_{int}$, 
where $L_{ret}$ is the loss for retrieval and $L_{int}$ for intention prediction.  $\lambda_1$ and $\lambda_2$ are hyper-parameters that work as scale factors.
\begin{equation}
    \mathcal{L}_{ret} = \sum_N \max( \rho_{neg}-(1-\rho_{pos})+\text{margin}).
\end{equation}
where $\rho_{neg}$ and $\rho_{pos}$ correspond to the cosine similarity of non-true (negative) stickers and true (positive) stickers. The margin is the rescaling margin.

\section{Experiment}
This section details the experimental settings and experimental results of our proposed \textbf{Int-RA} framework conducted on the created \textbf{StickerInt} dataset. We first present the experimental settings in Section \ref{Experimental Settings}. Then, we introduce various compared methods in Section \ref{Compared Methods}. We analyze the performance of our approach through the Main Results (\ref{Main Results}), Results on Other Datasets(\ref{Results on Other Datasets}), Ablation Study (\ref{Ablation Study}), Effect of Using Intention (\ref{Effect of Using Intention}), Effect of Number of Utterances (\ref{Effect of Number of Utterances}), and Effect of Different Attributes (\ref{Effect of Different Attributes}). Subsequently, we delve into the computational efficiency and scalability in Section \ref{Discussion on Computational Efficiency and Scalability}. Finally, to more intuitively demonstrate the performance of our method, we present some interactive cases in Section \ref{Case Study} and visualize the relations in Section \ref{Visualization}.

\subsection{Experimental Settings}
\label{Experimental Settings}
\textbf{Implement details.} 
We adopt Multi-lingual BERT \cite{devlin2018bert} as the text encoder to derive the textual representation. The CLIP pre-trained ViT model \cite{dosovitskiy2020image} is employed as the image encoder to derive the visual representation. We adopt Qwen-VL \cite{44-bai2023qwen} to generate sticker descriptions \footnote{In the preliminary experiment, we also try other MLLMs such as mini-GPT4 \cite{minigpt4} and Llava \cite{llava}. We found that Qwen-VL performed slightly better.}.
% The batch size is set to 4. 
We set the batch size to 4 and use Adam optimizer \cite{kingma2014adam} as our optimization algorithm. The learning rate is set to $1 \times 10^{-4}$. Both $\lambda_1$ and $\lambda_2$ are set to 1.

\textbf{Dataset.}
To evaluate the generalization ability of our method, we additionally conduct experiments on MOD \cite{4-fei2021towards} dataset, which 
includes bilingual (Chinese and English) dialogues and supports several tasks; we focus on the meme retrieval task. Each conversation has an average of 13.42 turns and 11.46 tokens per turn. As the official test set is unavailable, we report results on the validation set.

\textbf{Evaluation metrics.}
Two widely used evaluation metrics are applied in our experiments: mean of average precision (mAP), top N-precision (P@N).
mAP is a widely accepted criterion for assessing retrieval accuracy \cite{lin2014microsoft}.
P@N evaluates the precision of the top N predictions. Here, we mainly present the results for P@1, P@3, and P@5.
Notably, if the retrieved sticker matches the intention label of the ground truth sticker, we consider the result correct, as multiple stickers can serve as responses to the same conversation.
For the MOD dataset, we follow the original setting using recall at position k in n candidates R$_n$@k as an evaluation metric, which measures if the positive response is ranked in the top k positions of $n$ candidates.

\subsection{Compared Methods}
\label{Compared Methods}
To evaluate the performance of our model, we compare the proposed \textbf{Int-RA} with several baseline methods, including existing sticker retrieval methods, recent text-to-image retrieval approaches, and large language models-based methods.
\begin{itemize}
    \item \textit{Sticker retrieval methods}: \textbf{MOD} \cite{fei2021towards}, which leverages a unified generation network to produce multi-modal responses. \textbf{SRS} \cite{gao2020learning}, which learns sticker representations and utterance context in the multi-turn dialog. For a fair comparison, these two models are fine-tuned on our StickerInt dataset.
    \item \textit{Text-to-image retrieval methods}: \textbf{LGUR} \cite{shao2022learning} enhances feature granularity alignment via a transformer-based  framework. \textbf{IRRA} \cite{jiang2023cross} improves cross-modal alignment through implicit relation reasoning and similarity distribution matching. \textbf{PCME} \cite{chun2021probabilistic} introduces a probabilistic embedding model for cross-modal retrieval. \textbf{CLIP}, \textbf{Chinese-CLIP} \cite{radford2021learning}  and \textbf{ViLT} \cite{MOD-kim2021vilt} are encoder-based vision-language models, fine-tuned for sticker selection.
    \textbf{BLIP-2} \cite{MOD-li2023blip} and \textbf{FROMAGe} \cite{MOD-koh2023grounding} are encoder-decoder models. Given that FROMAGe is built on a large language model and pre-trained on large-scale multimodal dialogue data, we apply it directly to the task without additional fine-tuning.

    \item \textit{Large language models-based methods}: \textbf{Baichuan2}, \textbf{Llama3}, \textbf{ChatGLM3}, \textbf{Qwen2}, \textbf{Qwen-VL}, and \textbf{LLaVA}. 
    \textbf{Thanos} \cite{Model-lee2024thanos} infuses skill-of-mind reasoning into LLMs, generating both responses and conversational skills.
    \textbf{Ultron} \cite{Model-lee2024stark} is a multimodal conversational agent that incorporates social and contextual dynamics for image retrieval.
    \textbf{ChatIR} \cite{Model-levy2023chatting} is a dialogue-driven image retrieval system that interacts with users to iteratively refine queries and retrieve target images.
\end{itemize}

\begin{table}[!t]
\centering
\renewcommand\arraystretch{1.1}
\caption{Experimental results (\%) of various methods in sticker retrieval and text-to-image retrieval methods. \textbf{Bold} indicates that our method surpasses other models. We assert significance $^\ast$ if p-value < 0.05 under a t-test with the baseline models. w/o means without.}
\resizebox{\linewidth}{!}{
\begin{tabular}{ccccc}
\hline  
\textbf{Model}  &  \textbf{P@1} & \textbf{P@3} & \textbf{P@5} & \textbf{mAP} \\ 
\hline
%\textbf{ours}   &  \textbf{71.85} &\textbf{66.70} & \textbf{38.72} & \textbf{28.87}\\ 
\rowcolor{gray!20}  \multicolumn{5}{l}{\textit{Sticker retrieval methods}}\\
\multicolumn{1}{c}{SRS}  &1.30 &	3.25 &6.49&3.66  \\
\multicolumn{1}{c}{MOD} &5.84 	&9.74 &	14.29 &	9.72  \\
\hdashline
\rowcolor{gray!20}  \multicolumn{5}{l}{\textit{Text-to-image retrieval methods}}\\
\multicolumn{1}{c}{IRRA} &1.95 &	6.49&9.74 &6.10  \\
\multicolumn{1}{c}{PCME} & 3.90&11.69&19.48&11.40\\
\multicolumn{1}{c}{LGUR} & 9.09&11.69&15.58&12.10 \\
\multicolumn{1}{c}{Chinese-CLIP} &5.19&	12.34&20.13&12.89\\
\hdashline
\rowcolor{gray!20}  \multicolumn{5}{l}{\textit{Large language models-based methods}}\\
\multicolumn{1}{c}{LLaVA} & 6.49& 9.74& 14.94& 34.42 \\
\multicolumn{1}{c}{Qwen-VL} &9.09&24.68&30.52&33.41 
 \\
 Thanos &	3.90	&7.14&	17.53	&9.22\\

ChatIR &	9.09&24.68	&36.36	&25.63\\
\multicolumn{1}{c}{Qwen2} &13.64  &	25.32  &	27.92 	 &35.27 \\
\multicolumn{1}{c}{Baichuan2} &14.29&24.68&26.62 	&34.94  \\
Ultron 	&20.78	&20.78&	30.52&	25.21\\ 
\multicolumn{1}{c}{Llama3} & 15.58 &20.78&27.27&39.76  \\
\multicolumn{1}{c}{ChatGLM3} & 15.58 & 	25.32	 & 31.17 & 	36.29   \\
\hdashline
\textbf{Int-RA (ours)}&\textbf{18.18$^*$}& \textbf{37.66$^*$} & \textbf{40.91}$^*$&\textbf{53.37$^*$}\\
-SR scenario & 20.78	&39.61	&41.56	&56.79\\
-DR scenario  & 13.64&28.57&38.31&54.75\\
\hline
\end{tabular}
}
\label{tab2}
\end{table}

\subsection{Main Results}
\label{Main Results}
We examine the performance of our \textbf{Int-RA} framework in comparison with baselines across each evaluation metric and report the results in Table \ref{tab2}.
We also assess the significance of performance differences between the two runs using a two-tailed paired t-test, with strong significance at $\alpha=0.01$ denoted by $^*$.
It can be observed that our \textbf{Int-RA} consistently outperforms all baselines, demonstrating the effectiveness of the proposed \textbf{Int-RA} in sticker retrieval.
We can also notice an improvement in results as the value of N increases in Top N-precision, as more results can be utilized to expand the scope of potential matches with relevant labels.

\textbf{For sticker retrieval methods}, MOD and SRS perform significantly poorer compared to our \textbf{Int-RA}. This further highlights the efficacy of our approach in first predicting intention before conducting matching, emphasizing the crucial role of intention as a bridging component in the process.

\textbf{For text-to-image retrieval methods}, they prioritize capturing semantic relationships between text and image content. However, since they are not explicitly designed for sticker retrieval scenarios, exhibiting inferior performance compared to our framework.
In addition, an interesting observation is the overall superior performance of text-to-image retrieval methods compared to sticker retrieval baseline methods. This disparity can be attributed in part to the model design of SRS and MOD, where both models are devised to retrieve suitable stickers from a limited set of similar sticker candidates. 
Consequently, more attention is devoted to distinguishing local information among similar sticker expressions.
In contrast, our dataset encompasses a more diverse range of stickers in real-world conversations, requiring the model to analyze and recognize more complex visual features. This demonstrates the advantage of the new dataset and the proposed framework in this work.

\textbf{For large language models}, except for Ultron and ChatIR, we design a prompt that integrates the current session to generate the sticker description for each session, leveraging the in-context learning ability of LLMs \cite{LLM-li2023overprompt, LLM-xu2024unilog}. 
We then retrieve the appropriate sticker based on the generated sticker description and the combination of intent and sticker attributes using OpenAI’s LLM-based embedding model (text-embedding-ada-003) \cite{izacard2021unsupervised}.
We can observe that LLM models perform better than sticker retrieval and text-to-image retrieval methods. This superior performance is attributed to the extensive parameterization and complex network architecture of LLMs, which greatly enhance their ability to understand and generate complex language and image descriptions.
For P@1, Ultron performs the best, while for P@3 and P@5, ChatGLM3 achieves the highest performance, reaching 25.32\% and 31.17\%.
Compared to our IGSR method, all baseline models perform significantly worse. This further emphasizes the effectiveness of our approach in intent derivation, highlighting the critical role of intent as a bridging component in the process.

\textbf{Results of different scenarios.}  As illustrated in Figure~\ref{f-example}, our sticker retrieval task covers two typical usage scenarios: \textit{SR scenario}, where a sticker is used to supplement the current utterance, and \textit{DR scenario}, where a sticker is used to directly reply to a previous conversation without accompanying text. To assess the performance of our model in these two real-world settings, we conduct separate evaluations, with results summarized in Table~\ref{tab2}.
The comparison reveals that the model performs better under the SR scenario than the DR scenario, suggesting that retrieval is more effective when the sticker is used to complement explicit textual input. In contrast, the DR scenario presents greater challenges, as it requires the model to infer implicit intent from the broader dialogue context without direct linguistic cues. This discrepancy highlights the increased complexity of sticker retrieval in context-dependent, intention-driven interactions, and further emphasizes the necessity of intention modeling and commonsense reasoning in such cases.

% =============MOD===============
\begin{table}[!t]
\centering
\renewcommand\arraystretch{1.2}
\caption{Experimental results (\%) of various methods in MOD dataset. \textbf{Bold} indicates the best results, and \underline{underline} denotes the second-best. $^\dag$ represents the results from \cite{MOD-chen2024deconfounded}.}
\resizebox{0.9\linewidth}{!}{
\begin{tabular}{cccc}
\hline  
\textbf{Model}  &  \textbf{R$_{10}$@1} & \textbf{R$_{10}$@3} & \textbf{R$_{10}$@5}  \\ 
\hline
FROMAGe&9.83$^\dag$	&28.96$^\dag$	&49.85$^\dag$\\
ViLT  &	17.31$^\dag$	&40.00$^\dag$	&63.27$^\dag$\\
BLIP-2  &	17.64$^\dag$	&35.29$^\dag$&	47.18$^\dag$\\
CLIP &	\underline{21.90}$^\dag$  &	\textbf{47.72}$^\dag$&	66.04$^\dag$\\
MOD  &	\textbf{21.93}$^\dag$&	41.33$^\dag$	&\underline{70.78}$^\dag$\\
\textbf{Int-RA (ours)}	&\underline{21.90}&	\underline{44.50}	&\textbf{70.90}\\
\hline
\end{tabular}
}
\label{tab3}
\end{table}
% ============================

\subsection{Results on MOD Dataset}
\label{Results on Other Datasets}
We evaluate our proposed method Int-RA on a public benchmark dataset, MOD, to examine its generalization capability beyond our own StickerInt dataset. As shown in Table~\ref{tab3}, Int-RA achieves a R$_{10}$ @ 5 of 70. 90\%, slightly surpassing the best reported result by MOD. These results indicate that Int-RA, while not always the top-performing model across all metrics, can achieve performance on par with the best existing methods across diverse datasets, demonstrating strong generalization and robustness.

% =========ablation study=======
\begin{table}[!t]
\centering
\renewcommand\arraystretch{1.1}
\caption{Results of ablation study (\%). w/o means without.}
\resizebox{\linewidth}{!}{
\begin{tabular}{ccccc}
\hline  
\textbf{Model}  &  \textbf{P@1} & \textbf{P@3} & \textbf{P@5} & \textbf{mAP} \\ 
\hline

\textbf{Int-RA}&\textbf{18.18}& \textbf{37.66} & 40.91&\textbf{53.37}\\
\multicolumn{1}{l}{-w/o attribute}&17.53&35.06& \textbf{46.75}& 44.85 \\
\multicolumn{1}{l}{-w/o intention}&17.53 &35.71 & 41.56 &43.75\\
\multicolumn{1}{l}{-w/o knowledge}&18.18 &29.22 & 39.60 &39.36\\
\hline
\end{tabular}
}
\label{tab5}
\end{table}

\begin{table*}[!t]
\centering
\renewcommand\arraystretch{1.2}
\caption{Experimental results (\%) of different LLMs with or without intention.}
\resizebox{0.7\linewidth}{!}{
\begin{tabular}{ccccccccc}
\hline
\multicolumn{1}{c|}{\multirow{2}{*}{\textbf{Model}}} & \multicolumn{4}{c|}{\multirow{1}{*}{\textbf{Sticker Response}}}& \multicolumn{4}{c}{\multirow{1}{*}{\textbf{Utterance Response}}}\\ \cline{2-9}
\multicolumn{1}{c|}{} &\textbf{P@1} & \textbf{P@3} & \textbf{P@5} & \multicolumn{1}{c|}{\textbf{mAP}} &  \textbf{P@1} & \textbf{P@3} & \textbf{P@5} & \textbf{mAP}\\ 
\hline
\multicolumn{1}{l|}{\multirow{1}{*}{LLaVA}} &6.49 &9.74&14.94 &	\multicolumn{1}{c|}{\text{34.44}}&\text{16.23}	&\text{20.78}&	\text{29.87}	&\text{39.35}
  \\
\multicolumn{1}{r|}{-w/ intention} & \text{8.44}&	\text{25.97}	&\text{27.27}	&\multicolumn{1}{c|}{33.71}&	11.69&	19.48	&29.22	&35.34
  \\ \hdashline

\multicolumn{1}{l|}{\multirow{1}{*}{Qwen2}}  &\text{13.64}&	25.32&	27.92	&\multicolumn{1}{c|}{35.27}	&8.44	&20.13&	27.27	&35.48
 \\
\multicolumn{1}{r|}{-w/ intention}  &  9.09&	\text{25.97}	&\text{32.47}&	\multicolumn{1}{c|}{\text{37.39}}	&\text{9.09}	&\text{23.38}	&\text{28.57}	&\text{39.85}
  \\ \hdashline

\multicolumn{1}{l|}{\multirow{1}{*}{Baichuan2}} &\text{14.29} &	24.68 &	26.62 &	\multicolumn{1}{c|}{\text{34.94}} &	11.04	 &21.43	 &\text{28.57} &	33.11
 \\
\multicolumn{1}{r|}{-w/ intention}  &12.34&	\text{25.32}&	\text{28.57}&	\multicolumn{1}{c|}{31.52}	&\text{12.34}	&21.43	&27.92&	\text{36.17}
   \\ \hdashline
\multicolumn{1}{l|}{\multirow{1}{*}{ChatGLM3}} & \text{15.58}	&\text{25.32}&	\text{31.17}&	\multicolumn{1}{c|}{\text{36.29}}	&10.39&	22.73&	28.57&	\text{39.50}
 \\
\multicolumn{1}{r|}{-w/ intention}  & 12.99&	22.08	&25.97&	\multicolumn{1}{c|}{35.48}&	\text{12.34}	&\text{25.97}&	28.57&	33.84
    \\ \hdashline
\multicolumn{1}{l|}{\multirow{1}{*}{Llama3}} &\text{15.58} &	20.78	 &\text{27.27}	 &\multicolumn{1}{c|}{\text{39.76}} &	\text{12.99} &	\text{24.03}	 &\text{24.68} &	29.51

\\
\multicolumn{1}{r|}{-w/ intention}  &10.39 &	\text{22.73} &	25.97	 &\multicolumn{1}{c|}{33.50} &	9.74 &	18.83	 &24.03	 &\text{33.29}
 \\ \hdashline
\multicolumn{1}{l|}{\multirow{1}{*}{Qwen-VL}} &9.09	 &\text{24.68}	 &\text{30.52} &	\multicolumn{1}{c|}{33.41} &	\text{14.94}	 &21.43 &	25.97	 &\text{40.33}\\
\multicolumn{1}{r|}{-w/ intention}  &11.04& 20.13& 25.32  &\multicolumn{1}{c|}{35.50}&11.69& 25.97 &32.47 &37.69	
  \\ 
\hline
\end{tabular}
}
\label{LLM}
\end{table*}

\subsection{Ablation Study}
\label{Ablation Study}
We also conduct an ablation study on the use of knowledge and attributes. The evaluation results are shown in Table \ref{tab5}.
The performances of all ablation models are worse than those of the complete model under all metrics, which demonstrates the necessity of each component in our approach.
Note that the removal of attributes ("w/o attribute") results in considerable performance degradation, indicating that utilizing attributes can lead to better learning of sticker representation in different sticker properties.
Notably, by observing the performance of w/o attribute on P@5, we find that the impact of attributes is not as significant when a larger number of stickers are recalled.
In addition, the removal of intention tag ( "w/o intention" ) sharply degrades performance, which verifies the importance of knowledge in understanding conversation context. This demonstrates the crucial role of the intention tag in improving the accuracy and relevance of sticker retrieval tasks, highlighting the importance of considering users' intentions in retrieving appropriate stickers.

It is worth noting that the absence of the commonsense knowledge ("w/o knowledge") leads to a more significant decline in performance, with the mAP and P@3 scores decreasing by 14.01\% and 8.44\%, respectively.
This quantitative result underscores that the commonsense knowledge derived from relations like xIntent, xNeed, xWant, xEffect, and xReact is not just a performance booster but a key component for the intention predictor to make more interpretable and contextually grounded predictions.

\subsection{Effect of Using Intention}
\label{Effect of Using Intention}
To analyze the impact of intention, 
We introduce the intention into various LLMs as mentioned in the baseline. The results are shown in Table \ref{LLM}. "w/ intention" indicates that the large language model first predicts the intent of the response and then generates the reply based on the previous conversation context and the intent.
"Sticker Response" and "Utterance Response" indicate whether the model generates a description of the sticker or generates a response for the user. 
Overall, the sticker description approach performs better than the text response approach, indicating that generating a sticker description can more directly highlight the key points of the response, thus improving the retrieval of the corresponding sticker. 
Furthermore, we can also find that predicting the intention first and then generating the response is less effective than directly generating the response for most LLMs. 
That is to say, an incorrect initial prediction of intention can lead to inappropriate responses, ultimately reducing performance. This suggests that merely introducing intention does not guarantee improved model performance. In contrast, our approach enables the model to simultaneously perform sticker retrieval and learn response intent, resulting in superior performance.

\begin{figure}[!t]
  \centering
  \includegraphics[width=\linewidth]{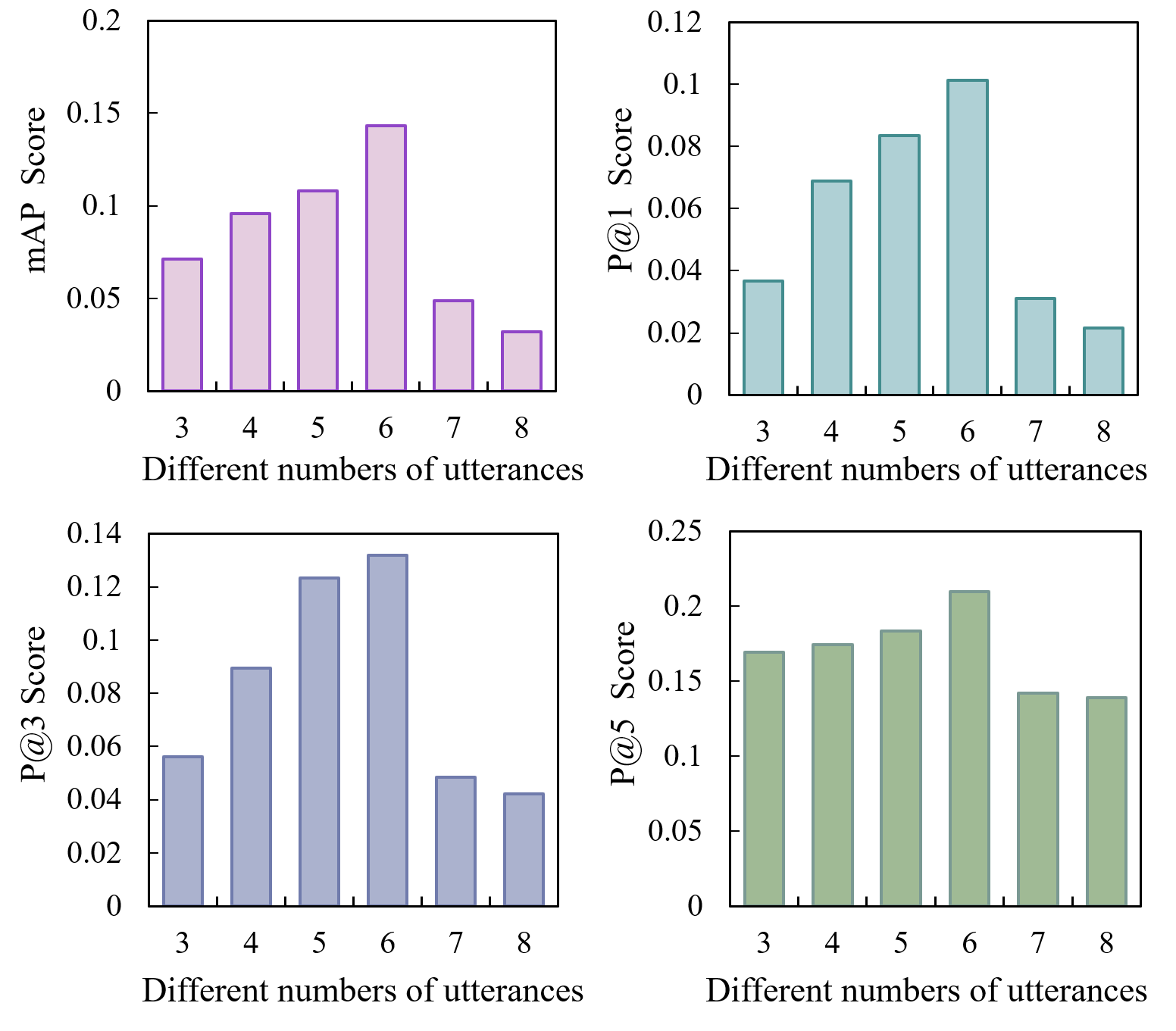}
  \caption{Performance of our approach on all metrics with different numbers of utterances.}
  \label{f5}
\end{figure}

\subsection{Effect of Number of Utterances}
\label{Effect of Number of Utterances}
To examine and analyze the impact of the number of utterances on the performance of our proposed \textbf{Int-RA } framework, we conduct experiments by varying the number from 2 to 10 and demonstrate the results in Figure \ref{f5}. 
We observe a similar trend across all evaluation metrics: mAP, P@1, P@3, and P@5. The results initially increase until the number of utterances reaches 6, after which they decrease as the number of utterances continues to increase.
One potential reason may explain this phenomenon. 
As the number of utterances increases from one to six, the model benefits from richer contextual information, which enhances its understanding of the conversation and improves retrieval performance. However, when the context becomes too long, utterances that appear much earlier than the sticker response may become less relevant, introducing noise and hindering accurate sticker selection. In our dataset, using six utterances offers the best balance between context richness and noise, leading to optimal performance.

\begin{table}[!t]
\centering
\renewcommand\arraystretch{1.4}
\caption{Experimental results (\%) of different attributes.  $\surd$ represents the used attribute. Ges., Pos., Face., and Ver. indicate gesture, posture, facial expression, and verbal, respectively.}
\resizebox{\linewidth}{!}{
\begin{tabular}{cccc|cccc}
\hline
\textbf{Ges.}  &\textbf{Pos.}  &\textbf{Face.}  &\textbf{Ver.}  &  \textbf{P@1} & \textbf{P@3} & \textbf{P@5} &\textbf{mAP}\\ 
\hline
% ges
$\surd$ && &  &9.09 &22.08&31.17&38.32
 \\
 % pos
&$\surd$ &  & &10.39&25.97 	&37.66&	49.72   \\
   % face
 & & $\surd$ & &9.09&27.27 &37.66 & 39.76 \\
 % ver
  & & &  $\surd$&12.99  &26.62  &37.01 &48.97\\ \hdashline
 % ges_pos
$\surd$ & $\surd$&  & &12.34 &22.73&36.36 &42.62\\
% ges_face
$\surd$  & & $\surd$ & &11.04 &35.06&36.36 &39.12 \\
% ges_ver
$\surd$  & &  & $\surd$ &10.39&	30.52&\textbf{42.86} &46.28\\
% pos_face
  & $\surd$& $\surd$ & &12.99 &29.22&42.21 & 	44.82
\\
% pos_ver
  & $\surd$&  & $\surd$&14.94 &27.92&33.77 &45.98\\
% face_Ver
  & & $\surd$ & $\surd$&13.64&25.32&37.66 	&49.95\\ \hdashline
% ges_pos_face
  $\surd$ & $\surd$& $\surd$ &&16.23 	&27.27 	&40.91 &47.54 \\
% get_pos_ver
  $\surd$ & $\surd$&  &$\surd$ &15.58  &29.22  &	40.26 &48.66\\
% pos_face_Ver
   & $\surd$& $\surd$ & $\surd$ &14.29 &25.97 	&38.31 &	46.00 \\ \hdashline
  $\surd$  & $\surd$& $\surd$& $\surd$  &  \textbf{18.18$^*$}& \textbf{37.66$^*$} & {40.91}&\textbf{53.37$^*$}\\
\hline
\end{tabular}
}

\label{attr_tab}
\end{table}

\subsection{Effect of Using Different Attributes}
\label{Effect of Different Attributes}
In the process of attribute-aware sticker representation, four visual attributes are utilized in our proposed \textbf{Int-RA} to represent the key expression information of the sticker. This section examines the effectiveness of different attributes. The results of various scenarios of attribute combinations are shown in Table \ref{attr_tab}.
It can be seen that the performance overall improves with the increase in the number of attributes used, and using all the attributes achieves the best performance. Using only one attribute significantly degrades the performance, indicating that the visual information can not be learned sufficiently from a single perspective. That is, relying solely on this single attribute is insufficient for capturing the full expressive range necessary for accurate sticker representation. Further, incorporating multiple attributes provides richer information, thereby leading to improved performance. This concludes that a more holistic approach that combines multiple attributes to understand the visual information of stickers is essential for optimal performance.

\begin{figure*}[!t]
  \centering
  \includegraphics[width=\linewidth]{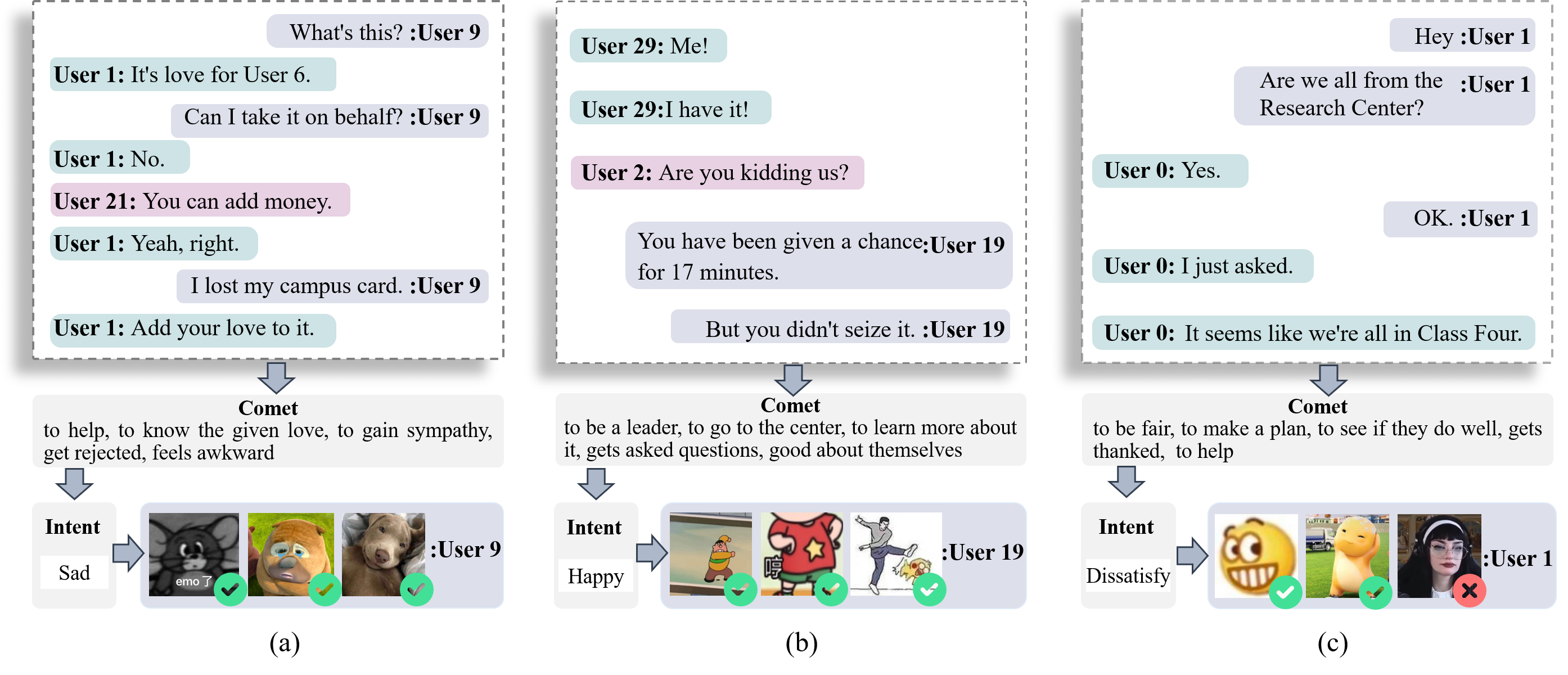}
  \caption{Examples of conversation context, commonsense information, predicted intent, and top-3 stickers retrieved by our method.}
  \label{f7}
\end{figure*}

\subsection{Discussion on Computational Efficiency and Scalability}
\label{Discussion on Computational Efficiency and Scalability}
To further explore the practicality of the proposed Int-RA framework, we examine its computational efficiency and scalability. 
The multi-stage architecture of Int-RA, comprising commonsense-based intention prediction, attribute-aware sticker description generation via MLLM, and relation-aware cross-modal matching, contributes to increased inference time.
We analyze the overall computational cost of Int-RA in terms of three primary components: intention prediction time $t_{int}$, attribute-aware representation generation time $t_{attr}$, and relation-aware matching time $t_{match}$. The total inference time can be approximated as:
$t=t_{int}+t_{attr}+t_{match}=O(n_r \cdot L)+O(k \cdot d)+O(k\cdot d^2)$. where $n_r$ is the number of COMET relations, $L$ is the input sequence length for commonsense inference, $k$ is the number of candidate stickers, and $d$ is the embedding dimension for both visual and textual representations. In our experiment, Int-RA achieves an average response time of approximately 1.25 seconds on a single V100 GPU, which is acceptable for most real-time human-computer interaction applications. 

Additionally, as demonstrated in our ablation studies (Section~\ref{Ablation Study}), each component meaningfully contributes to final performance. Nevertheless, simplified variants of the model still retain competitive performance, offering flexibility in efficiency-performance trade-offs. 

\subsection{Case Study}
\label{Case Study}
Several interactive cases retrieved by our approach are provided in Figure \ref{f7}. These conversation samples suggest that our pipeline framework holds the capacity to provide sticker-incorporated expressive communication.
From examples (a) and (b), we can observe that our approach tends to favor stickers with similar actions and facial expressions, with the characteristics of emoji stickers often being manifested through detailed features such as gestures and facial expressions. This demonstrates the effectiveness of using visual attributes to enhance the learning of stickers.
However, in example (c), the final prediction is incorrect, likely due to the diverse styles of stickers, which remains a challenge in the current state of sticker recognition. 

Furthermore, in example (a), User 9's response is not a direct reply to the preceding utterance but rather addresses the emotional expression in the historical conversation \textit{"I lost my campus card"}. This shows the difficulty of understanding conversation context in multi-user conversation, resulting in the increased challenge of sticker retrieval. Consequently, in future research, user information can be considered to further improve the performance of sticker retrieval.

To investigate the interpretability of the intention predictor, we analyze how commonsense knowledge contributes to intention inference. In example (a), the generated commonsense inferences include phrases such as “get rejected” and “feel awkward”, which strongly suggest a feeling of sadness. The predicted intention aligns with this emotional tone, enabling the model to select a sticker that appropriately conveys the underlying sentiment. This example illustrates how the intention predictor functions as a transparent and semantically grounded bridge between dialogue understanding and sticker selection. By leveraging external knowledge to infer user intent, the model enhances both retrieval accuracy and human interpretability.

\begin{figure}[!t]
  \centering
  \includegraphics[width=0.9\linewidth]{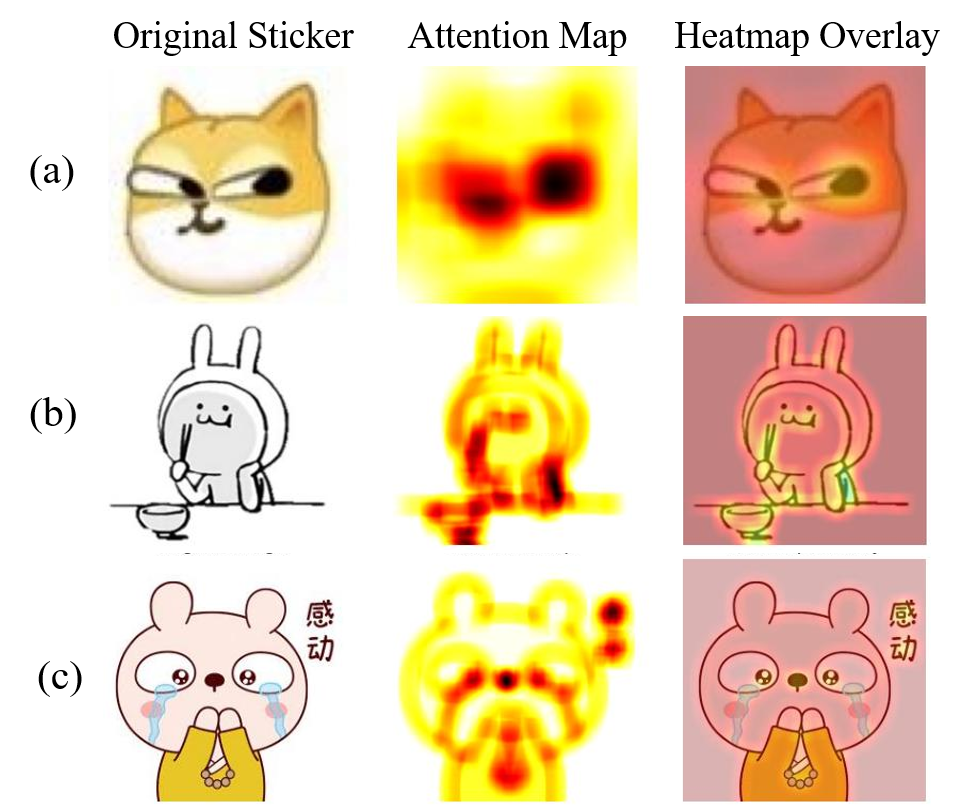}
  \caption{Visualization of the relation score on stickers. The darker the color, the higher the relation score.}
  \label{attention}
\end{figure}

\subsection{Visualization}
\label{Visualization}
To analyze how our \textbf{Int-RA} learns the important information about stickers, we visualize the relation score $\mathcal{M}$ (Equation \ref{att_eqa}) of three stickers in Figure \ref{attention}.
For example (a), where the character appears neutral. This indicates that the representation of this sticker heavily relies on this facial expression. Our \textbf{Int-RA} can effectively catch the important information in the sticker by the relation score placement on the character's face.
Moreover, the relation score also attends to the character's gestures. For instance, in Case (b), where the character is depicted as holding chopsticks with one hand and supporting the face with the other, we observe attention focused on his hand, suggesting that our \textbf{Int-RA} learns key points of body language.
Furthermore, considering that the relation score comprehensively considers four properties of stickers, as illustrated in Case (c), we observe that our \textbf{Int-RA} pays attention to both facial expressions and gestures simultaneously, thereby learning accurate visual information about the sticker.

\section{Conclusion and Future Work}

We create a new dataset for sticker retrieval, called \textbf{StickerInt}. Unlike previous studies that view stickers merely as a supplement to the current utterance, our new dataset can cover two real-world scenarios of using stickers in online conversation: using stickers to reply to previous conversations or supplement our words. 
Based on the new dataset, we propose \textbf{Int-RA}, a framework for sticker retrieval in conversation. In which, the intention information is leveraged in the learning of conversation context. Further, we devise four novel visual attributes, i.e., \textit{gesture}, \textit{posture}, \textit{facial expression}, and \textit{verbal}, to improve the learning of stickers. Based on this, a relation-aware sticker selector is explored to retrieve the sticker for the conversation.
Extensive experiments conducted on our \textbf{StickerInt} dataset and other public datasets demonstrate that our proposed approach achieves outstanding performance in sticker retrieval.

In the future, we will focus on developing advanced models that can better handle the diversity of sticker styles in real-world conversations and improve the accuracy of sticker retrieval in multi-user, multi-turn dialogues. Furthermore, we plan to explore personalizing sticker recommendations and exploring cross-cultural differences in sticker usage to enhance the versatility and applicability of our models across various user groups and scenarios.

\section{Limitations}
The limitations of this work are mainly twofold. 
Firstly, stickers have diverse styles ( e.g., cartoon, animal, etc.) in real-world conversations, which might affect the performance of the sticker retrieval task. 
Additionally, real-world conversations often involve multiple users engaging in multi-turn conversations. In such scenarios, our method may not fully capture the complexities of interactions among multiple users.
Future research could focus on addressing these limitations by exploring more sophisticated models or incorporating additional contextual information to improve the performance of the sticker retrieval task.

\section*{Acknowledgments}
This work was partially supported by 
the National Natural Science Foundation of China (62176076), 
the Natural Science Foundation of Guangdong (2023A1515012922), 
the Shenzhen Foundational Research Funding (JCYJ20220818102415032), the Guangdong Provincial Key Laboratory of Novel Security Intelligence Technologies (2022B1212010005), Hong Kong RGC GRF No. 14206324, CUHK direct grant No. 4055209, and CUHK Knowledge Transfer Project Fund No. KPF23GWP20.

\bibliographystyle{IEEEtran}
\bibliography{IEEEabrv,references}

\newpage

\vspace{11pt}

\vfill

\end{document}